\documentclass[traditabstract, letter]{aa}
\usepackage{longtable,lscape,amssymb}
\usepackage{graphicx}
\usepackage{epsfig}
\usepackage[authoryear]{natbib}

\bibpunct{(}{)}{;}{a}{}{,} 

\bibpunct{(}{)}{;}{a}{}{,} 

\begin{document}
 

\newcommand{\Ha}{{H$\alpha$}}
\newcommand{\Hb}{{H$\beta$}}
\newcommand{\Hg}{{H$\gamma$}}
\newcommand{\Hd}{{H$\delta$}}
\newcommand{\Hep}{{H$\epsilon$}}

\newcommand{\Ht}{H$_2$}
\newcommand{\CIV}{C\,{\sc iv}}
\newcommand{\CII}{C\,{\sc ii}}
\newcommand{\MgII}{Mg\,{\sc ii}}
\newcommand{\NV}{N\,{\sc v}}

\newcommand{\pab}{{Pa$\beta$}}
\newcommand{\pag}{{Pa$\gamma$}}
\newcommand{\pad}{{Pa$\delta$}}
\newcommand{\brg}{{Br$\gamma$}}
\newcommand{\brd}{{Br$\delta$}}

\newcommand{\Ll}{{$L_{\rm line}$}}
\newcommand{\LHa}{{$L_{\rm H\alpha}$}}
\newcommand{\Lacc}{{$L_{\rm acc}$}}
\newcommand{\Macc}{{$\dot{M}_{\rm acc}$}}
\newcommand{\BJobs}{{BJ$_{\rm obs}$}}
\newcommand{\BJintr}{{BJ$_{\rm intr}$}}

\newcommand{\Msun}{{$M_{\odot}$}}
\newcommand{\Lsun}{{$L_{\odot}$}}
\newcommand{\Rsun}{{$R_{\odot}$}}

\newcommand{\Mjup}{{$M_{jup}$}}

\newcommand{\Mstar}{{$M_{\star}$}}
\newcommand{\Lstar}{{$L_{\star}$}}
\newcommand{\Rstar}{{$R_{\star}$}}
\newcommand{\Teff}{{$T_{\rm eff}$}}

\newcommand{\Av}{{$A_{\rm V}$}}

\newcommand{\Mdisk}{{$M_{\rm disc}$}}
\newcommand{\Ldisk}{{$L_{\rm disc}$}}
\newcommand{\Rin}{{$R_{\rm in}$}}

\newcommand{\SO}{{$\sigma$-Ori}}

\newcommand{\gq}{{GQ\,Lup}}
\newcommand{\gqa}{{GQ\,Lup\,A}}
\newcommand{\gqb}{{GQ\,Lup\,B}}
\newcommand{\gqc}{2MASS\,J15491331}
\newcommand{\parlup}{{Par-Lup\,3-4}}


\title{2MASS\,J15491331-3539118: a new low-mass wide companion of the GQ Lup system \thanks{
Based on observations  collected at the European Southern Observatory at Paranal, 
under program 103.C-0200(A), and archive data from 074.C-0037(A) and 082.C-0390(A).}
}

\author{
       J. M. Alcal\'a\inst{1}
  \and F. Z. Majidi\inst{2,3}
  \and S. Desidera\inst{3} 
  \and A. Frasca\inst{4}
  \and C. F. Manara\inst{5}
  \and E. Rigliaco\inst{3}
  \and R. Gratton\inst{3}
  \and M. Bonnefoy\inst{6} 
  \and E. Covino\inst{1}
  \and G. Chauvin\inst{6}
  \and R. Claudi\inst{3}
  \and V. D'Orazi\inst{3}
  \and M. Langlois\inst{7}
  \and C. Lazzoni\inst{3}
  \and D. Mesa\inst{3}
  \and J.E. Schlieder\inst{8}
  \and A. Vigan\inst{9}
}

\offprints{J.M. Alcal\'a}
\mail{juan.alcala@inaf.it}

\institute{ 
       INAF-Osservatorio Astronomico di Capodimonte, via Moiariello 16, 80131 Napoli, Italy
  \and Dipartimento di Fisica, dell'Universit\'a di Roma Sapienza, P.le A. Moro, 5, Roma I-00185, Italy 
  \and INAF-Osservatorio Astronomico di Padova, vicolo dell'Osservatorio 5, 35122 Padova, Italy
  \and INAF-Osservatorio Astrofisico di Catania, via S. Sofia, 78, 95123 Catania, Italy
  \and European Southern Observatory, Karl-Schwarzschild-Str. 2, D-85748 Garching bei Munchen, Germany
  \and Universit\'e Grenoble Alpes, CNRS, IPAG, 38000 Grenoble, France
  \and CRAL, UMR 5574, CNRS, Universit\'e de Lyon, Ecole Normale Sup\'erieure de Lyon, 46 All\'ee d'Italie, F-69364 Lyon Cedex 07, France
  \and Exoplanets and Stellar Astrophysics Laboratory, Code 667, NASA Goddard Space Flight Center, Greenbelt, MD 20771, USA
  \and Aix Marseille Universit\'e, CNRS, CNES, LAM, Marseille, France
}

\date{Received ; accepted  }

\abstract{ 
Substellar companions at wide separation around stars hosting planets or brown dwarfs (BDs) yet close 
enough for their formation in the circumstellar disc are of special interest. In this letter we report 
the discovery of a wide (projected separation $\sim$16\farcs0, or 2400\,AU, and position angle 114.61$^\circ$) 
companion of the \gqa-B system, most likely gravitationally bound to it. 
A VLT/X-Shooter spectrum shows that this star, 2MASS\,J15491331-3539118, is a bonafide low-mass 
($\sim$0.15\,\Msun) young stellar object (YSO) with stellar and accretion/ejection properties typical of 
Lupus YSOs of similar mass, and with kinematics consistent with that of the \gqa-B system. A possible scenario 
for the formation of  the triple system is that \gqa ~and 2MASS\,J15491331-3539118 formed by 
fragmentation of a turbulent core in the Lup~I filament, while \gqb, the BD companion of \gqa ~at 0\farcs7, 
formed {\em in situ} by the fragmentation of the circumprimary disc. The recent discoveries that stars form 
along cloud filaments would favour the scenario of turbulent fragmentation for the formation of \gqa ~and 
2MASS\,J15491331-3539118.
}


\titlerunning{New companion to GQ\,Lup}
\authorrunning{Alcal\'a et al.}
\maketitle

\section{Introduction} 
\label{intro}

It is now well established that most stars form in binaries or higher-order multiple systems 
\citep[][and references therein]{monin07}. 
Substellar wide companions around stars that host planets or brown dwarfs (BDs) at 
separation close enough for formation in the circumstellar disc are of special interest. 
The dynamical environment of young binaries and multiple systems may have an important impact 
on disc evolution \citep[][]{kurtovic18, manara19} and therefore on planet formation. 
During the early evolution phase, the individual discs of substellar companions, including those 
in the planetary-mass regime, accrete additional material from the gas-rich parent disc, and therefore their 
discs are more massive and their accretion rates higher than objects of similar mass formed in 
isolation. This means that disc masses and accretion rates of these very low-mass companions are expected to 
be independent of the mass of the central substellar object, and higher than predicted from 
the \Macc$\propto$\Mstar$^2$ scaling relation for more massive young stellar objects (YSOs). 
These arguments were used by \citet[][]{stamatellos15} to explain the very high levels of accretion 
observed in substellar and planetary-mass companions at wide orbits in some T Tauri stars 
\citep[][]{zhou14}. 
 
The results of ALMA observations of wide multiple systems \citep[][]{williams14, kurtovic18, manara19} 
have provided important information on the disc morphologies and geometry of these systems, revealing misalignments
of the disc and orbital angular momentum vectors, which provide important clues about the formation mechanisms 
of these systems. Likewise, the discovery of ultra-wide pairs \citep[UWPs; 1000--6000\,AU;][]{joncour17} 
is shedding light on the formation mechanisms at separation scales of the pristine star forming cores.
 
The Gaia DR2 catalogue \citep[][]{gaiadr2} offers the opportunity to study the kinematic and astrometric 
properties of large samples of stars on large sky areas, allowing the selection of comoving YSO candidates  
in star forming regions with similar parallaxes and kinematics to those of bonafide YSOs. 
As part of an investigation (Majidi et al., in prep.) aimed at characterising companions to SPHERE targets 
hosting giant planets and discs, we searched for wide companions associated to SHINE SPHERE-GTO targets 
\citep[][]{chauvin17} in the Gaia DR2, finding several tens of candidates. We focused on a few objects 
around stars of high specific interest due to the presence of planetary or BD companions and spatially 
resolved circumstellar discs. \gq , a $\sim$1\Msun \citep[][]{alcala17, macgregor17}~classical T Tauri star, 
has both a spatially resolved circumstellar disc \citep[][]{macgregor17} 
and a BD companion  \citep[\gqb;][]{neuhauser05} with a mass about 20-40\,\Mjup ~at a separation of 0\farcs70. 
This BD companion was found to be actively accreting  \citep[][]{zhou14}, suggesting the presence of a disc 
around it, although more recent observations show that the accretion is non-steady \citep[][]{wu17}.
 
In this letter we report the discovery of a wide companion of the \gq ~system that is most likely gravitationally 
bound,  making it a very interesting triple system. In Section~\ref{obs}, the target 
selection and observations with the VLT/X-Shooter spectrograph are described. In Section~\ref{results}, 
we characterise the new companion in terms of its stellar physical and accretion/ejection 
parameters. In Section~\ref{discussion}, we discuss the nature of \gq ~as a triple system and the implications 
on its possible formation mechanism. 
 
\section{Target selection and observations} 
\label{obs}
 
We searched the Gaia DR2 catalogue for candidates within 5\,arcmin of the selected targets with 
similar parallaxes and proper motions. We identified a companion candidate\footnote{By {\it wide companion candidate} 
we refer to the object spatially proximate to the \gq ~system, with a projected separation of a few tens of arcsec,
but not necessarily gravitationally bound to the system.} at 16\farcs0 of \gq ~and position angle 114.61$^\circ$. 
This candidate is identified with 2MASS\,J15491331-3539118 and is slightly fainter 
(by 0.2-0.5\,mag) in the $J$, $H,$ and $K$ bands (14.85, 14.08 and 13.82, respectively) than \gqb. 
Its mass was then expected to be slightly lower ($\sim$20\,\Mjup ~for 2\,Myr age) than \gqb. From its near-infrared (NIR) colours 
and position on a $J-H$ versus $H-K$ diagram, the object appears moderately reddened (\Av$\sim$0.5--1\,mag) and 
does not show strong NIR excess. However, whether the discovered source is gravitationally bound to the \gq ~system 
remains inconclusive. Therefore, although it could be referred to as GQ\,Lup\,C, for caution we refer 
to it as \gqc, and for the reasons discussed in Appendix~\ref{Appstelpar}, we adopt the distance of 
152($\pm$1)\,pc from the Gaia DR2 as for \gqa ~ \citep[][]{gaiadr2}.

\begin{figure}[h]
\resizebox{0.97\hsize}{!}{\includegraphics[bb=-10 20 740 520]{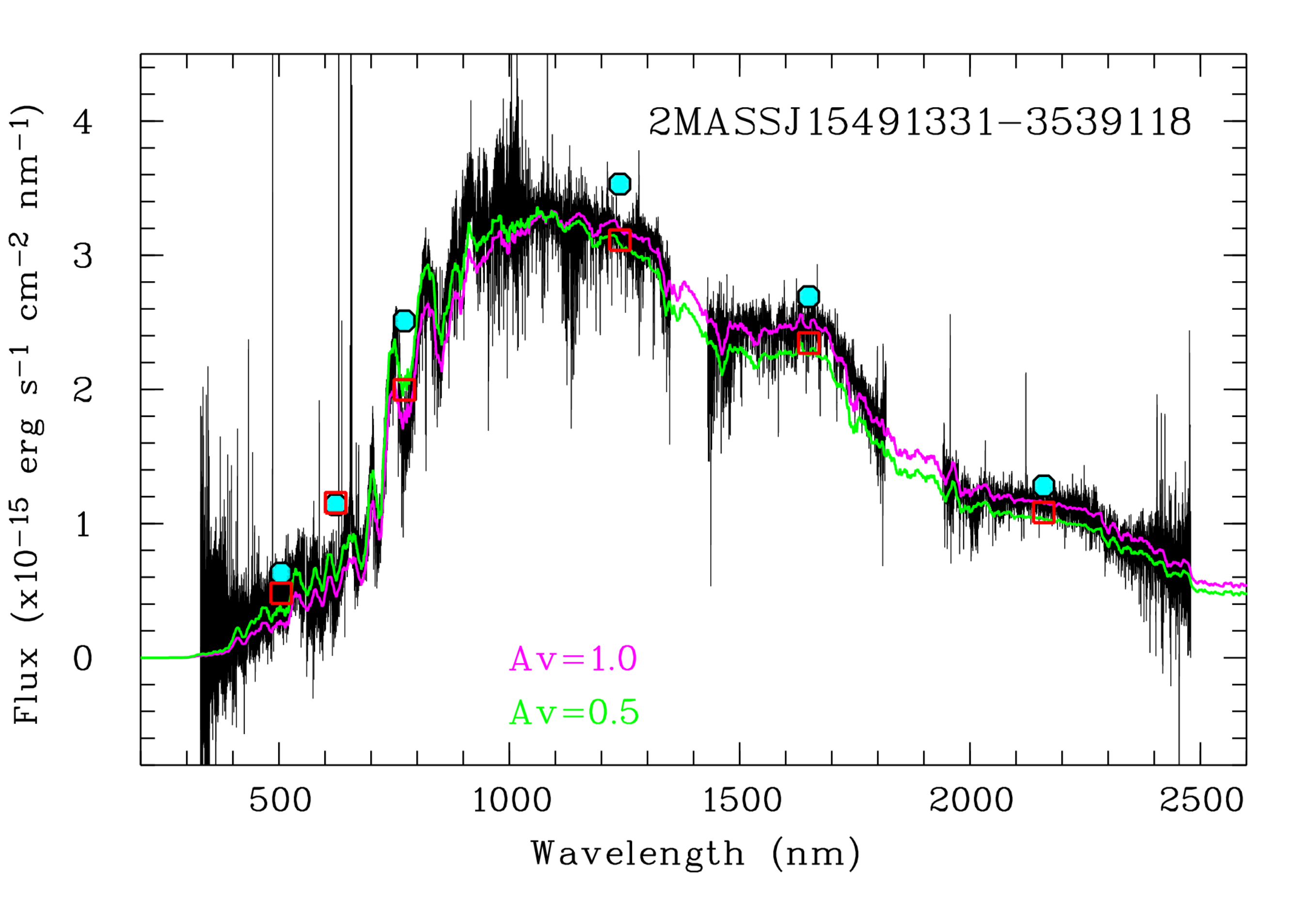}}
\caption{ X-Shooter spectrum of \gqc ~(in black) compared with photometric fluxes (light blue dots) 
         from the Gaia DR2 and 2MASS catalogues. Synthetic fluxes derived from the observed spectrum 
         are shown as red squares. The lines show a BT-Settl model of the same effective 
         temperature as the target, for the two reddening values indicated in the labels. The model 
         has been normalised to the flux of \gqc ~at 830\,nm.
         \label{sed}}
\end{figure}

We investigated \gqc ~using the VLT/X-Shooter spectrograph \citep{vernet11}. 
Service mode observations were conducted during the night of 28 May 2019.  
We adopted the same observational and data processing strategies as in previous investigations of YSOs 
in Lupus \citep[see][and Appendix~\ref{obstrategy} for the details]{alcala14,alcala17}.
Figure~\ref{sed} shows the flux distribution of the reduced X-Shooter spectrum of \gqc ~compared with the 
photometric fluxes available in the Gaia DR2 and 2MASS. The flux calibration of the spectrum yields results 
consistent with the photometric fluxes within the flux calibration uncertainties of 15--20\%.

\section{Results}
\label{results}

\subsection{The spectrum of \gqc} 
The spectrum of \gqc ~is typical of a late-type YSO showing a wide (W\Ha(10\%)$\sim$510\,km/s) 
and strong (EW$\sim$100\AA) \Ha ~emission, and the Li~I (6708\AA) absorption line (See Figure.~\ref{spec_Ha} 
in Appendix~\ref{AppMacc}), as well as numerous permitted and forbidden emission lines, the latter 
being of comparable intensity to the former. This behaviour may be explained by a contrast effect 
possibly due to obscuration of the star and of the accretion flows by the inner disc regions; it  is 
similar to the one observed in the spectrum of \parlup, a subluminous, edge-on YSO in the Lupus~III cloud 
\citep[][]{alcala14,alcala17}, and suggests that we are viewing \gqc ~at a high inclination angle 
close to edge-on. The Balmer decrement is also consistent with such a geometry (See Appendix~\ref{Baldecr}).
Many Balmer lines and several He\,{\sc i} and Ca\,{\sc ii} lines and the \pab ~ and \brg ~lines 
are observed in emission. We were able to measure the flux of the Balmer lines up to H\,15, as well as 
of several other emission lines (see Appendix~\ref{AppMacc}). 
Despite a low S/N in the UV, U-excess continuum emission in \gqc ~is also evident (see Figure~\ref{UV_fit}
in Appendix~\ref{AppMacc}). All these results confirm that the object is a bonafide YSO with high mass 
accretion and ejection rates. In the following we use the X-Shooter spectrum combined with the 
Gaia DR2 data to characterise the object.

\subsection{Stellar parameters} 
\label{stellpar}
The physical stellar parameters given in Table~\ref{stelpar1} were derived using the same methods applied
for our previous X-Shooter surveys in star forming regions \citep[][]{alcala14, alcala17, manara17}. 
The details are given in Appendix~\ref{Appstelpar}. The same methods yield a very similar extinction value
\citep[\Av$=0.70$\,mag,][]{alcala17} for \gqa.  A fit of a BT-Settl model spectrum \citep[][]{allard12} of 
the same \Teff ~as the object suggests a moderate extinction of \Av$\sim$1.0\,mag for \gqc
\footnote{We note that \citet[][]{petrus19} show that extinction may be overestimated by the BT-Settl models.} 
(see Figure~\ref{sed}), which is consistent with the value derived using our methods.

Unexpectedly, the resulting spectral type M4 indicates that \gqc ~is a star, and not a BD. We note that the luminosity 
derived from the bolometric flux of the star (see Appendix~\ref{Appstelpar}) is more than an order of magnitude lower 
than that of Lupus YSOs of similar \Teff, again suggesting subluminosity due to obscuration of the inner disc 
regions. This may also explain the faint NIR magnitudes in comparison with Lupus objects of similar spectral type. 
The stellar radius, \Rstar, derived from \Lstar ~ and \Teff ~ is also underestimated.

\setlength{\tabcolsep}{2.5pt}
\begin{table}
\caption[ ]{\label{stelpar1} Physical stellar parameters of \gqc } 
\begin{tabular}{c|c|c|c|c|c}
\hline \hline

SpT          &  \Av              & \Teff           &  \Lstar        &  \Rstar       & \Mstar   \\
                &  (mag)         & (K)             &  (\Lsun)       &  (\Rsun)      &(\Msun)            \\
\hline     
               &                 &                  &               &               &            \\

M4$\pm$0.5       &   1.0$\pm$0.5 &     3190$\pm$100  & 0.004$\pm$0.001 & 0.21$\pm$0.06 &  0.15$\pm$0.05    \\

\hline
\end{tabular}
\end{table}

\setlength{\tabcolsep}{2.5pt}
\begin{table}
\caption[ ]{\label{stelpar2} Parameters derived from ROTFIT} 
\begin{tabular}{c|c|c|c|c|c}
\hline \hline

\Teff     &  $\log{g}$      &  \Rstar$^\dagger$  & \Lstar$^\dagger$   & $RV$          &  $v \sin{i}$     \\
(K)      &                  &  (\Rsun)     &  (\Lsun)   &  (km/s) & (km/s)         \\
\hline     
         &                  &       &     &  &         \\

3230$\pm$101 &  3.74$\pm$0.23 &  0.87$\pm$0.20 &   0.07$\pm$0.02  &  $-$2.0$\pm$2.8  &  13.0$\pm$6.0  \\

\hline
\end{tabular}
\tablefoot{~\\
$\dagger$ : values derived from $\log{g}$ \& \Mstar.}
\end{table}

The ROTFIT code \citep[][]{frasca15} applied to the X-Shooter spectrum yields a \Teff ~which is 
in excellent agreement with the value derived from the spectral type and surface gravity typical of 
low-mass PMS stars. The results are provided in Table~\ref{stelpar2}. The radial velocity ($RV)$ is 
consistent, within errors, with the values in the literature for the \gq ~system. ROTFIT operates on
spectral segments normalised to unity, and therefore a measure of \Av ~ is not provided by the code.

An estimate of \Rstar\ that is  independent of \Lstar\  can be obtained using the ROTFIT $\log g$ value and 
assuming the mass of 0.15\Msun, which is not affected by subluminosity because the PMS tracks run almost 
vertically at the \Teff ~values close to that of \gqc. When doing so, we obtain a radius of 0.87\Rsun ~for 
the star and hence a luminosity that is higher by a factor of $\sim$17 than the estimated \Lstar 
~based on the bolometric flux. This places  \gqc ~ on the Hertzsprung-Russell (HR) diagram in good agreement 
with other Lupus members, with an estimated age of $\sim$2\,Myr based on the \citet[][]{baraffe15} theoretical 
isochrones (See Figure~\ref{HRD} in Appendix~\ref{Appstelpar}), and consistent within errors with the age of 
the \gq ~system of $\sim$1\,Myr \citep[][]{frasca17}.

\subsection{Mass accretion and ejection}
\label{accretion}

The analysis to derive the accretion luminosity, \Lacc, which measures the Balmer 
continuum excess emission with respect to the photospheric one, is described in Appendix~\ref{AppMacc}. 
We estimated  $\log$(\Lacc/\Lsun)$=-3.54$. We also derived \Lacc ~ from the luminosity, \Ll , of a 
number of Balmer and other emission lines using the \Ll--\Lacc ~relationships \citep[][]{alcala17}. 
From this analysis we 
calculated $\log$\Lacc$=-3.74\pm0.20$, which is in good agreement with the value derived from the Balmer 
continuum fit (see Appendix~\ref{AppMacc} for details). 
However, we note  that this is about one order of magnitude lower than the typical \Lacc ~of Lupus YSOs 
of similar mass; the reason is again that the emission from the accretion funnels is obscured by the 
edge-on inner-disc regions. Therefore, as in other cases of subluminous YSOs \citep[][]{alcala14,alcala17},
we use the luminosity of the [O\,{\sc i}] $\lambda$6300\,\AA ~forbidden line to correct for these effects, 
deriving $\log$\Lacc$=-2.85\pm0.20$ (see Appendix~\ref{AppMacc} for details). 
The correction factor is very similar to the one on \Lstar, which is reasonable as the accretion funnels 
should be obscured by the inner disc to a similar extent to the stellar photosphere. 
The corrected accretion luminosity can be converted into a mass accretion rate using Eq.~\ref{EqMacc} and 
the ROTFIT corrected \Rstar ~value, yielding  $\log$\Macc$=-9.50$ or \Macc$=3.3\times10^{-10}$\,\Msun yr$^{-1}$,
which is within the typical range of values for Lupus YSOs of comparable mass.

The mass loss rate was estimated following \citet[][see Appendix~\ref{AppMacc} for details]{natta14}.
The value we derive is M$_{loss} \sim$7.0$\times$10$^{-12}$\,\Msun/yr, similar to the values for very 
low-mass YSOs, and implying a M$_{loss}$/\Macc ~ratio of $\sim$0.02 in the expected range for 
magneto-centrifugal jet launch models (0.01 $<$ M$_{loss}$/\Macc $<$ 0.5) and studies of accreting 
YSOs \citep[][and references therein]{cabrit09, whelan14, nisini18}.

\subsection{Kinematics}
\label{kinematics}
 
The radial velocity estimates in the literature for \gqa ~are $-$3.6$\pm$1.3\,km/s \citep[][]{frasca17} and
$-$2.8$\pm$0.2\,km/s \cite[][]{schwarz16}. We also retrieved ten spectra taken with 
HARPS (Program ID 074.C-0037(A) and 082.C-0390(A))   from the ESO
archive. The RVs drawn from the instrument pipeline yield an 
average of $-$3.3\,km/s with rms scatter of 0.8\,km/s (dominated by activity-induced RV jitter) over a 
baseline of 1470\,days. The ROTFIT RV measurement of $-2.0\pm$2.8\,km/s for \gqc ~is consistent, 
within errors, with these values. Likewise, the Gaia DR2 proper motion for \gqc 
~($\mu_{\alpha}\cos{\delta}=-14.81\pm0.97$\,mas \& $\mu_{\delta}=-21.95\pm0.65$\,mas) is consistent within 
$\sim$1$\sigma$ with that of \gqa ~($\mu_{\alpha}\cos{\delta}=-14.26\pm0.10$\,mas \& $\mu_{\delta}=-23.60\pm0.07$\,mas). 
Therefore, the kinematics of \gqc ~is highly consistent with that of the \gq ~system. 

\section{Discussion and conclusions}
\label{discussion}

Above, we demonstrate that \gqc ~is a bonafide low-mass YSO in the Lupus~I cloud, 
with stellar, accretion, and ejection properties typical of Lupus YSOs of similar mass, and kinematics 
consistent with that of the \gqa-B system. At the distance of the \gq ~system, 16\farcs0 corresponds to a separation 
of about 2430\,AU. Multiples with such separations are not expected to be very common \citep[][]{baron18}, 
but if \gqc ~is physically bound to \gqa ~it would be another example of a young low-mass wide-bound system 
\citep[see][for other cases]{petrus19}, and also an interesting case for further studies 
with high-resolution millimetre observations with ALMA 
\citep[see][for ALMA observations in similar systems]{williams14, kurtovic18, manara19}. 

The separation of 2430\,AU of \gqc ~ is larger than that of the wide pairs so far studied with ALMA, but 
it is in the low tail of the separation distribution (1000--60000\,AU) for UWPs recently 
detected by \citet[][]{joncour17} in the Taurus star forming region. As noted in that work, 
the UWPs with separations of $<$5000\,AU have been found to be physical wide binaries \citep[][]{kraushill09}. 
This would favour the conclusion that \gqa-B-\gqc is likely a gravitationally bound system.
It is difficult to firmly establish whether \gqc ~is gravitationally bound to \gq ~with our data alone. 
The escape velocity at such a large separation is less than 1 km/s, which is comparable with the expected 
rms value due to turbulence and well below the accuracy of the current estimates of the relative motions. 
However, \gqa ~and \gqc ~are likely still gravitationally bound. In fact, at a velocity of 1\,km/s, a star
travels 1\,pc in 1\,Myr, roughly the estimated age of the \gq ~system \citep[][]{frasca17}; 
1\,pc is $\sim$22\,arcmin at the distance of \gq, which is two orders of magnitude more than the separation 
of the \gq ~system components. Moreover, the probability of a chance projection is very small, and can  
be ruled out based on the statistical grounds presented in Appendix~\ref{chanceprojection}.

The inclination $i=60.2^{\circ}$ of the circumprimary disc \citep[][]{macgregor17,ansdell18} and 
the edge-on geometry of the \gqc ~disc might in principle be an indication that the angular momentum 
vectors are not aligned. Misalignments have been observed in other multiple 
systems with components at closer separations than \gqc ~\citep[][]{brinch16, kurtovic18, williams14, 
jensen14, manara19}. These studies provide important clues for the mechanisms leading to misalignments 
during the early evolutionary phase, and discuss some possible scenarios, such as for example fragmentation of a turbulent 
core where the directions of the angular momentum vectors change. This is supported by hydrodynamical 
simulations \citep[][]{bate18} that indicate misaligned discs are mainly due to fragmentation in turbulent 
environments and stellar capture; alternatively, the dynamical interactions of three or more protostars 
may lead to random changes of the orbital axes during the early Class~0-I phases. Also, the environmental 
conditions in which the discs are formed may affect the alignments with a variety of disc 
orientations being the result of gas accreted from the parent cloud with a different angular momentum. 
With our data alone it is not possible to establish the relative geometries of the \gqa-B -\gqc ~discs. 
However, evidence for the alignment of the system will be presented in a follow-up paper 
(Lazzoni et al. in prep.). We also note that the orbital analysis by \citet[][]{schwarz16} suggests 
a $\sim$60$^\circ$ inclined orbit for \gqb, that is, very similar to that of the circumprimary disc.  
 
A firm conclusion on the formation mechanism of the \gqa-B-\gqc ~system would be premature. Given its high mass-accretion 
rate, the {\em in situ} formation of \gqb ~by circumprimary disc fragmentation \citep[][]{stamatellos15} cannot 
be ruled out \citep[][]{wu17, macgregor17}, although the presence of \gqc ~might have had an impact on its orbit 
due to dynamical interaction. \citet[][]{joncour17} concluded that the young UWPs may be pristine imprints of 
their spatial configuration at birth resulting from sequential fragmentation of the natal molecular core. 
Therefore, a likely scenario for the formation of the \gqa-B-\gqc ~system may be that \gqa ~and \gqc ~formed by 
fragmentation of a turbulent core, with \gqb ~formed {\em in situ} by the fragmentation of the circumprimary 
disc. The recent result that stars form along filaments 
\citep[e.g.][and references therein; see also Appendix~\ref{chanceprojection}]{benedettini18}
would favour the scenario of turbulent fragmentation for the formation of \gqa ~and \gqc.
Further high-resolution interferometric observations of wide multiple systems with discs in millimetre bands will 
shed light on their formation mechanisms.

\begin{acknowledgements}
We thank the anonymous referee for his/her comments and suggestions.
This work has been supported by the project PRIN-INAF 2016 The Cradle of Life - GENESIS-SKA 
"General Conditions in Early Planetary Systems for the rise of life with SKA", and by the 
project PRIN-INAF-MAIN-STREAM 2017 
"Protoplanetary disks seen through the eyes of new-generation instruments". 
This project has received funding from the European Union's Horizon 2020 research and 
innovation programme under the Marie Sklodowska-Curie grant agreement No 823823 (DUSTBUSTERS).
This work was partly supported by the Deutsche Forschungs-Gemeinschaft (DFG, German Research 
Foundation) - Ref no. FOR 2634/1 TE 1024/1-1.
This research made use of the SIMBAD database, operated at the CDS (Strasbourg, France). 
This work has made use of data from the European Space Agency (ESA) mission Gaia 
(https://www.cosmos.esa.int/gaia), processed by the Gaia Data Processing and Analysis 
Consortium (DPAC, https://www.cosmos.esa.int/web/gaia/dpac/ consortium). 
Funding for the DPAC has been provided by national institutions, in particular the 
institutions participating in the Gaia Multilateral Agreement. 
\end{acknowledgements}

\begin{appendix}
\normalsize

\section{Observational strategy and data processing}
\label{obstrategy}
The X-Shooter science exposure of \gq ~ consisted in a two-cycle (A-B-B-A) nodding 
mode observation for a total on-source exposure of 1.2\,hr using the 1\farcs0 slits; as part of 
the same observing block, an additional short exposure (of $\sim$10\% the science exposure) was 
performed in stare mode using the wide slits of 5\farcs0 immediately before the science observation 
for slit-loss correction purposes. Such a strategy allows us to achieve a precision in absolute 
flux calibration on the order of 15--20\%. The slit was aligned at parallactic angle.

Data reduction was performed using the X-Shooter pipeline version 3.3.5 in the v2.9.1 reflex 
environment \citep[][]{freudling13}. The steps of the data processing are the same as those described 
in \citet[][]{alcala14,alcala17}. In particular, the absolute flux calibration was done using the 
wide-slit short exposure spectrum to correct for slit losses.

\section{Stellar physical parameters}
\label{Appstelpar}

Spectral type, SpT, was calculated using the spectral indices by \citet{riddick07} and the H2O-K2 
index from \citet{rojasayala12} for the NIR. Extinction is estimated by comparison of the VIS spectrum with that of spectral templates best matching 
the spectrum of \gqc. All the templates have a low extinction ($A_{\rm V} <$ 0.5) and are taken from
\citet[][and references therein]{manara17}. The templates were then artificially reddened by 
$A_{\rm V}$ in the range 0--4.0\,mag in steps of 0.25\,mag, until the best match to the object spectrum was found. 
The effective temperature, \Teff, was estimated using the derived spectral type and the relationship between 
spectral type and \Teff ~by \citet[][]{HH14}. Other calibrations \citep[e.g.][]{luhman03, pecmam13}  provide
very similar \Teff ~values.

The Gaia DR2 parallax (5.49$\pm$0.46\,mas) for \gqc ~yields a distance d$=$182($\pm$15)\,pc, while for most of the 
YSOs in Lupus the distance is 159\,pc \citep[see appendix in][]{alcala19}. For \gqa, ~the parallax 
(6.59$\pm$0.05\,mas) yields a distance d$=$152($\pm$1)\,pc. The error on the measurement for \gqc ~is rather large, 
with the estimated-excess-noise of 3\,mas and astrometric-excess-noise-sig of 59, making it rather 
unreliable. This is probably due to the edge-on geometry of the object which may affect the photocentre 
measurement. We note that other objects, such as Sz\,102 in Lupus, RY\,Tau in Taurus \citep[see][]{garufi19}, and 
R\,CrA \citep[][]{mesa19}, with similar edge-on geometries to that of \gqc ~or extended circumstellar structures 
suffer from the same DR2 problem. 
As concluded in Sect. 4, the proximity of \gqc ~to the \gq ~system by a chance projection 
is very unlikely, with the stars most likely being gravitationally bound. Thus, it is reasonable to adopt 
the distance of \gqa ~for \gqc, which in any case is different by just $\sim$2$\sigma$.

\begin{figure}[h]
\resizebox{1.0\hsize}{!}{\includegraphics[bb=20 20 780 550]{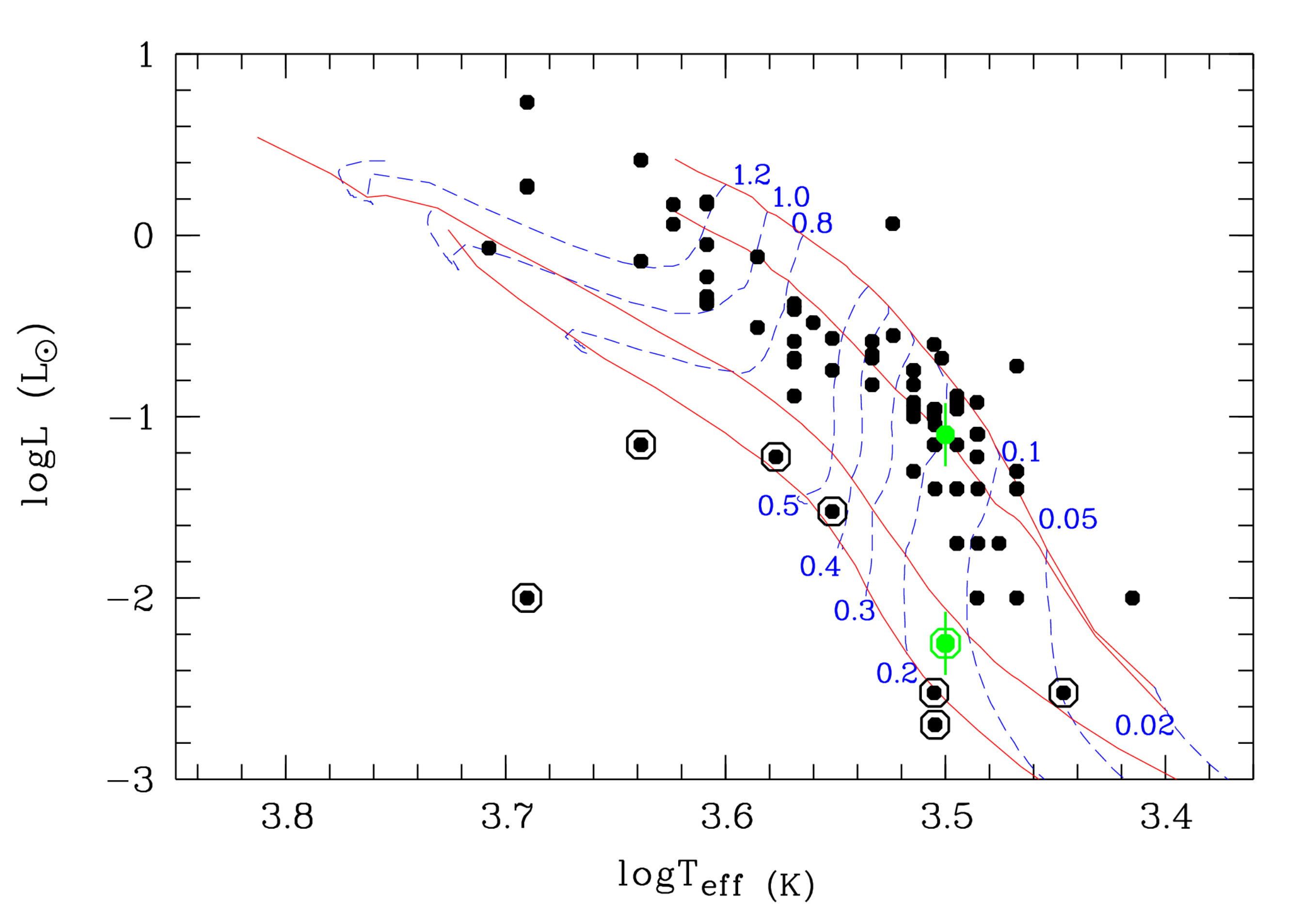}}
\caption{HR diagramme of the Lupus sample (black dots) with  \gqc ~overplotted in green. 
        The PMS evolutionary tracks of \citet[][]{baraffe15} are overplotted with the dashed 
        blue lines, while isochrones (1\,Myr, 3\,Myr, 30\,Myr and 10\,Gyr) are the red lines.
        Subluminous objects are overplotted with encircled dots (\gqc ~in green).
        The green non-encircled dot is \gqc ~when luminosity is corrected by disc obscuration. 
         \label{HRD}}
\end{figure}

The bolometric luminosity, \Lstar, was derived by integration of the X-Shooter spectrum after correction for 
extinction using the \citet[][]{cardelli89} extinction law and adopting the distance $d=$152\,pc; to 
include in the integration spectral regions not covered by X-Shooter, the spectrum was extrapolated both in 
the blue and red using the BTSettl model of the same \Teff ~as \gqc. The contribution of this extrapolation to 
the bolometric flux represents less than 10\% of the total bolometric flux. 
The stellar radius, \Rstar ~was derived from \Teff ~and \Lstar ~using the Stefan-Boltzmann law. 
We note that \Rstar ~is underestimated for subluminous objects and, as discussed in the main text, \gqc ~is 
subluminous with respect to other Lupus members. This can be seen in Figure~\ref{HRD} where the position of 
the star on the HR diagram is shown before and after applying the correction using the luminosity-independent 
\Rstar ~value based on the $\log{g}$ measurement. The latter has been derived by applying the code ROTFIT 
\citep[][]{frasca15} to the X-Shooter spectrum. 
The code finds the best photospheric template spectrum that reproduces the target spectrum by minimising the
$\chi^2$ of the difference between the observed and template spectra in specific spectral segments. 
The spectral segments are normalised to unity, and therefore a measure of \Av ~ is not provided by the code.
The spectral intervals analysed with ROTFIT contain features that are sensitive to the effective temperature and/or 
$\log{g}$, such as the Na\,{\sc i} doublet at $\lambda \approx 819$\,nm and the K\,{\sc i} doublet at
$\lambda \approx 766-770$\,nm. The code also allows us to measure the $v \sin{i}$ ~by $\chi^2$ minimisation applied 
to spectral segments devoid of broad lines.
The results produced by ROTFIT are \Teff$=$3230$\pm$101\,K, a surface gravity $\log{g}=3.74\pm0.23$, radial velocity 
$RV=-2.0\pm2.8$\,km/s and projected rotational velocity $v \sin{i}=13.0\pm6.0$\,km/s.

Finally, the stellar mass was estimated by comparison with the low-mass PMS evolutionary tracks by 
\citet[][ See Figure~\ref{HRD}]{baraffe15}. 

\section{Measurements of mass accretion and ejection rate}
\label{AppMacc}

\begin{figure}[h]
\resizebox{1.0\hsize}{!}{\includegraphics[bb=0 20 750 520]{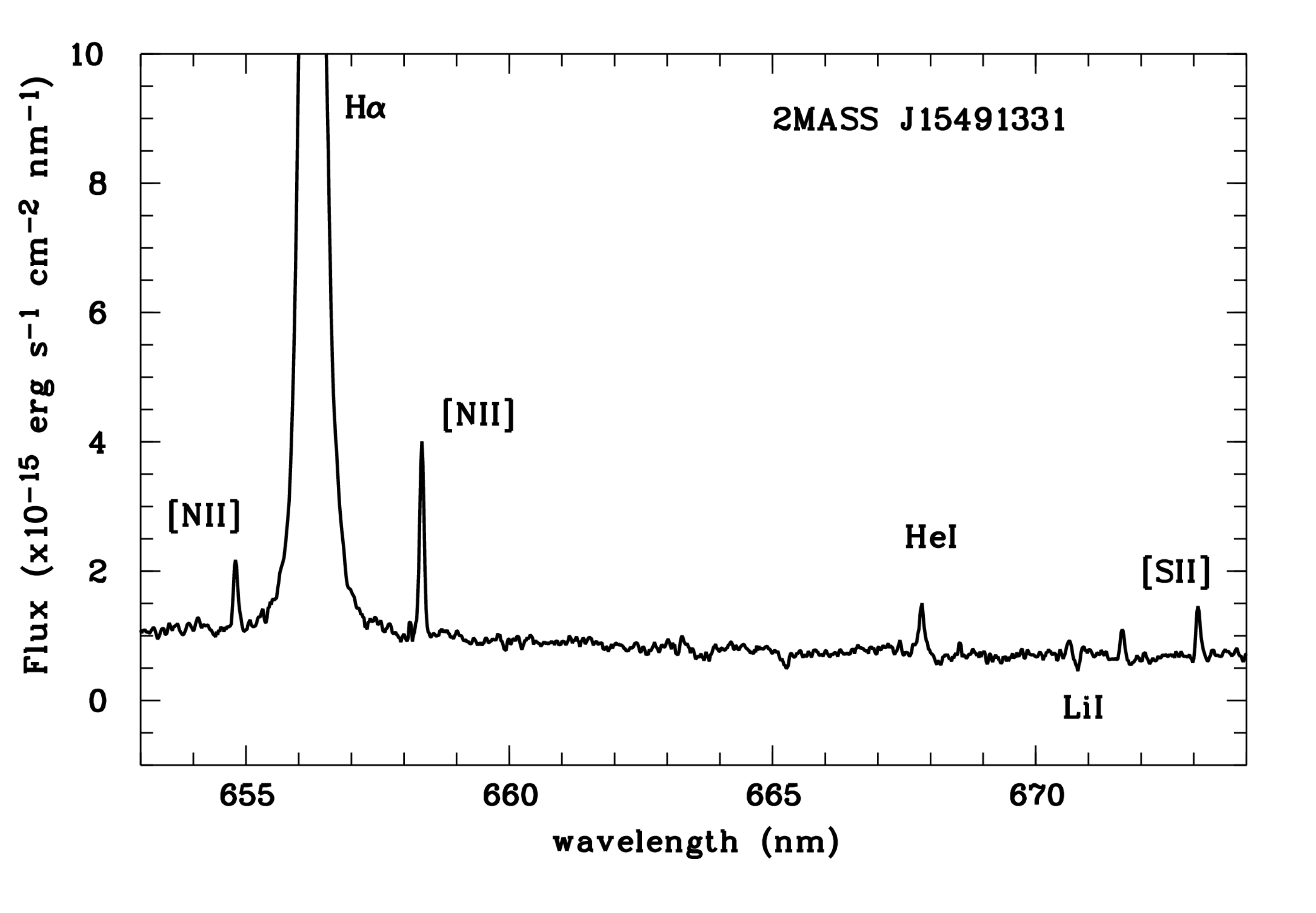}}
\caption{Portion of the X-Shooter spectrum of \gqc ~around the \Ha ~line. The He\,{\sc i} $\lambda$6678\AA 
        ~and the forbidden lines of [N\,{\sc ii}] and [S\,{\sc ii}] and the Li\,{\sc i} $\lambda$6708\AA 
        ~are labelled. 
         \label{spec_Ha}}
\end{figure}

\begin{figure}[h]
\resizebox{1.0\hsize}{!}{\includegraphics[bb=0 -5 110 50]{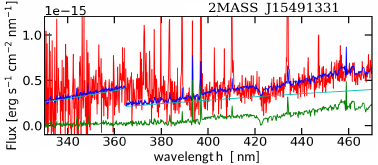}}
\caption{Ultraviolet spectrum of \gqc ~in the region of the Balmer jump (red line). The spectrum of 
        the adopted non-accreting template SO\,797 \citep[][]{manara13} is overplotted 
        with a green line. The continuum emission from the slab is shown by the cyan 
        continuous line. The best fit with the emission  predicted from the slab model 
        added to the template is given by the blue line.
         \label{UV_fit}}
\end{figure}

The spectrum of \gqc ~shows strong emission lines, both permitted and forbidden. The \Ha ~line is
intense (EW$\approx$100\AA) and broad ($\Delta$V$>$500\,km/s), the satellite [N\,{\sc ii}] 
forbidden lines are seen in emission (see Figure~\ref{spec_Ha}), and  UV-excess emission is 
evident in the spectrum (Figure~\ref{UV_fit}), suggesting high mass accretion and ejection rates.

To derive the accretion luminosity, \Lacc, we followed the methods described in \citet[][]{alcala14, 
alcala17} and \citet{manara17}. Briefly, the spectrum of \gqc ~was fitted as the sum of the photospheric 
spectrum of a non-accreting template and the emission of a slab of hydrogen; the accretion luminosity is 
given by the luminosity emitted by the slab. The stellar and accretion parameters are self-consistently 
derived by finding the best fit among a grid of slab models and using the continuum UV-excess emission 
and the broad wavelength range covered by the X-Shooter spectra (330\,nm -- 2500\,nm) to constrain 
both the spectral type of the target and the interstellar extinction toward it. Figure~\ref{UV_fit} shows 
the fit. From our analysis we estimate that in general the uncertainty on \Lacc ~is $\sim$0.25\,dex 
\citep[see][]{alcala14, alcala17, manara17}.
The best fit yields \Lacc$=2.55\times10^{-4}$\,\Lsun ~or $\log$\Lacc$=-3.6$ and simultaneously the stellar 
parameters SpT=M4.5, \Av$=0.5$\,mag, \Teff$=$3200\,K, and \Lstar$=$0.0043\,Lsun. All the results from the fit 
are highly consistent with those derived in Appendix~\ref{Appstelpar}, within the errors.

\begin{figure}[h]
\resizebox{1.0\hsize}{!}{\includegraphics[bb=10 20 740 520]{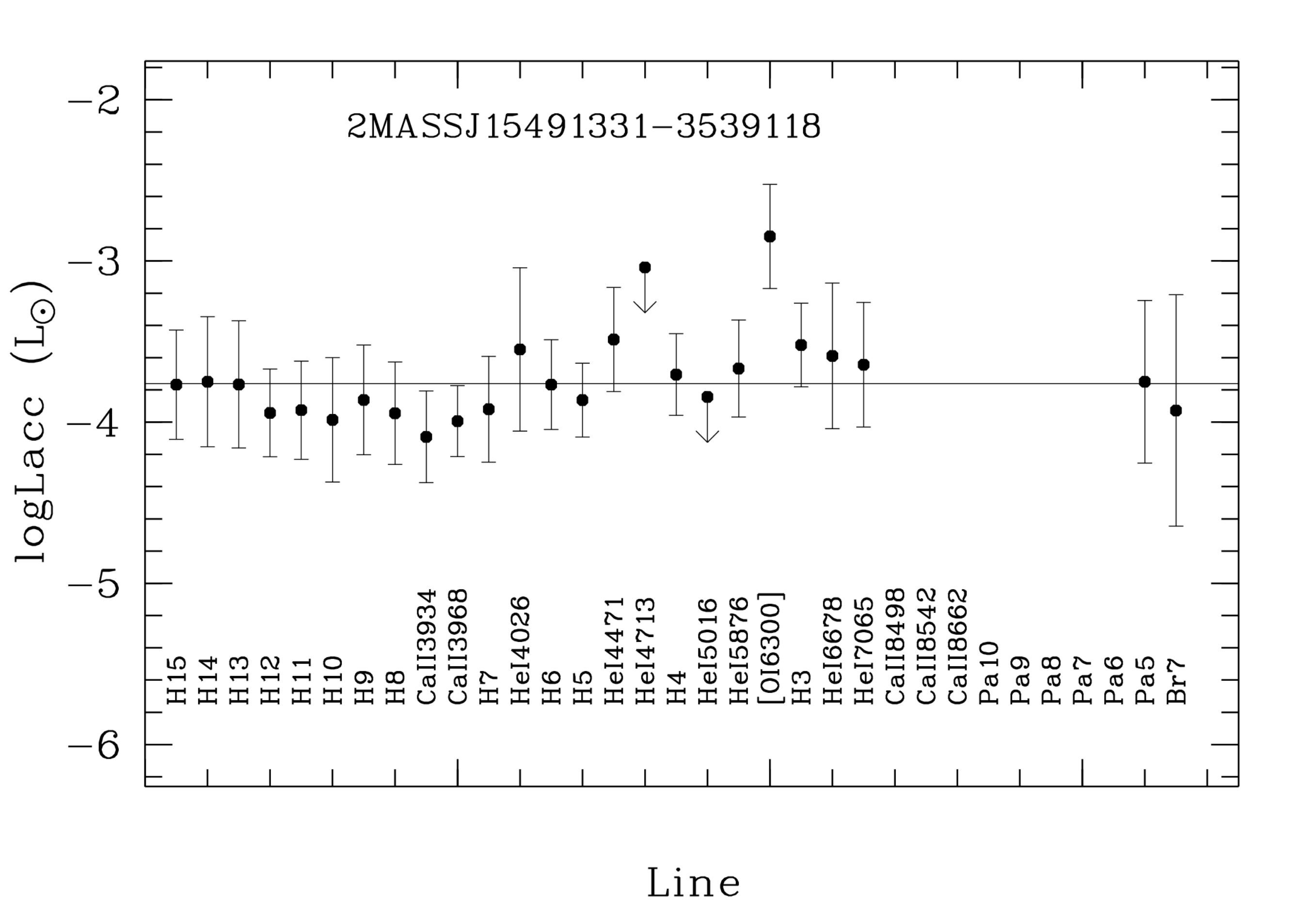}}
\caption{Accretion luminosity as a function of different line diagnostics. The different diagnostics
        are labelled and sorted by increasing wavelength. The horizontal line corresponds to the
        average value of \Lacc.
         \label{Lacc_lines}}
\end{figure}

We also used the luminosity, \Ll, of several permitted emission lines and \Ll--\Lacc ~relationships
\citep[][]{alcala17} to derive \Lacc. The \Ll ~of Balmer lines up to H15, and that of several  
He\,{\sc i} and Ca\,{\sc ii} permitted lines, were computed by measuring the line fluxes on the X-Shooter 
spectrum, correcting them by extinction (\Av$=$1\,mag) and adopting the Gaia DR2 distance of \gqa 
~(See discussion in Section~\ref{Appstelpar}).
Figure~\ref{Lacc_lines} shows \Lacc ~as a function of the diagnostics used. Here, \Lacc ~is consistent within 
errors among the different lines, except for the  [O\,{\sc i}] forbidden line. Excluding the latter, the 
average $\langle$\Lacc$\rangle$ ~yields $\log\langle$\Lacc$\rangle=-3.76$ with a dispersion of 0.20\,dex.

The  [O\,{\sc i}] line, which is also correlated with \Lacc ~\citep[][]{natta14, nisini18}, was deliberately 
included in our analysis because it originates quite far from the star and is free of obscuration from the 
inner disc. This is why \Lacc ~from this line is about one order of magnitude higher than the values estimated from 
the permitted lines. The other permitted lines form in the accretion columns in the inner disc regions and 
hence suffer obscuration by the almost edge-on disc, by a similar amount as the stellar luminosity. 
Thus, in principle we could estimate a correction of \Lacc  ~based on  \Lacc ~derived from the [O\,{\sc i}] line. 
A correction of 0.9\,dex in $\log$\Lacc ~can be calculated, and the corrected $\log$\Lacc ~is -2.84. 
This accretion luminosity was converted into mass accretion rate, \Macc, using the relation

{\setlength{\mathindent}{0pt}
\begin{equation}
\label{EqMacc}
\dot{M}_{acc} = ( 1 - \frac{R_{\star}}{R_{\rm in}} )^{-1} ~ \frac{L_{acc} R_{\star}}{G M_{\star}}
 \approx 1.25 ~ \frac{L_{acc} R_{\star}}{G M_{\star}} 
,\end{equation}
}

\noindent
where $R_{\star}$ and $M_{\star}$ are the radius and mass of \gqc, respectively, and $R_{\rm in}$ is the 
inner-disc radius \citep{gullbring98, hart98}. The latter corresponds to the distance from the star at 
which the disc is truncated -- due to the stellar magnetosphere -- and from which the disc gas is accreted, 
channelled by the magnetic field lines. It has been found that $R_{\rm in}$ ranges from 3$R_{\star}$ to 
10$R_{\star}$ \citep{johns-krull07}. For consistency with our previous studies 
\citep[e.g.][and references therein]{alcala14, alcala17, manara17}, here we assumed $R_{\rm in}$ to be 
$5\,R_\star$. We used the ROTFIT corrected \Rstar ~(see main text).

Besides the [O\,{\sc i}] optical lines, several other forbidden lines are identified in the VIS and UVB arms 
of the spectrum ([O\,{\sc ii}], [S\,{\sc ii}], [N\,{\sc ii}], see Figure~\ref{spec_Ha}). None of these lines show evidence of the high-velocity component (HVC) indicative of fast moving jets, but only show the low-velocity 
component (LVC) that is known to trace slow-moving winds (either thermally or magnetically driven).  Following 
\citet[][]{natta14}, a rough estimate of the mass loss through the winds traced by the LVC can be obtained 
assuming a simple spherical geometry for the emitting gas, just dividing the mass of the emitting region 
by a timescale $\tau$ = L/v where L and v are measured along the flow direction.  

In particular, we use the [O\,{\sc i}] $\lambda$6300\,\AA ~line; that from the flux-calibrated spectrum has a 
luminosity L$_{\rm [O\,I]}$=3.65$\times$10$^{-7}$\,\Lsun ~(assuming a distance of 152\,pc) or a factor 1.4 higher 
if the distance is 182\,pc. From Eq.(6) in \citet[][]{natta14} we find that M$_{gas}$=5.11$\times$10$^{-12}$\,\Msun. 
Moreover, we derive the volume occupied by the gas, and assuming spherical geometry, the linear dimension of the 
region emitting the observed  [O\,{\sc i}]. We find L$\sim$0.15\,AU, and assuming a peak velocity of $\sim$1~km/s, 
this translates into a typical timescale of $\tau \sim$0.8 years. The mass loss rate we obtain is therefore 
M$_{loss} \sim$7.0$\times$10$^{-12}$\,\Msun/yr (at 152 pc, or M$_{loss} \sim$ 8.9$\times$10$^{-12}$\,\Msun/yr 
at 182\,pc). 

\section{Balmer decrements}
\label{Baldecr}

\begin{figure}[h]
\resizebox{1.0\hsize}{!}{\includegraphics[bb=10 20 740 520]{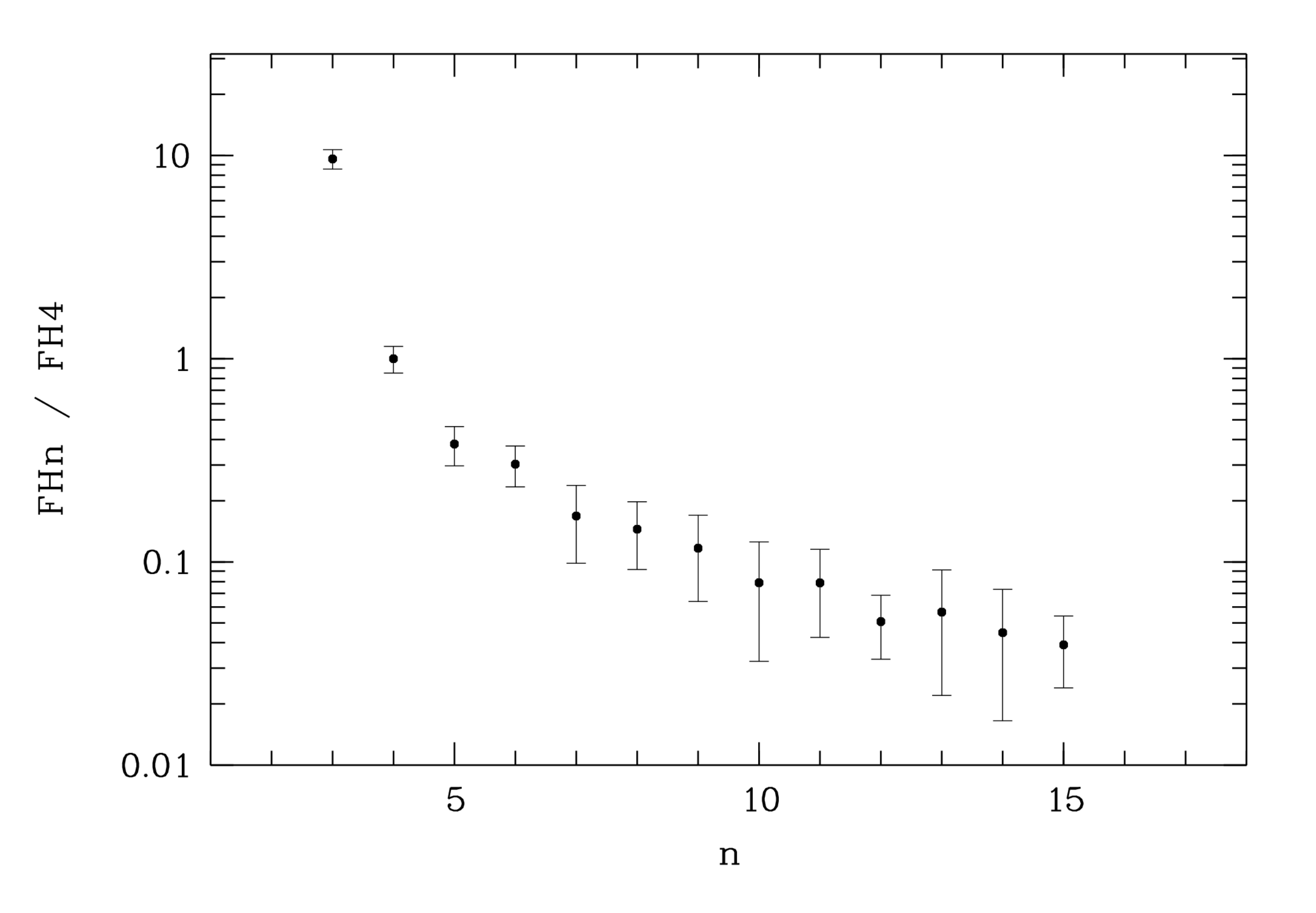}}
\caption{Balmer decrements as a function of the quantum number. The H$\beta$ line
        has been taken as reference for the analysis. 
         \label{balmerdecr}}
\end{figure}

We also investigated the Balmer decrements in \gqc. The results are shown in Figure~\ref{balmerdecr}.
We decided to use \Hb  ~(or H4) as reference for the same reasons as discussed in \citet[][]{antoniucci17}.
The high \Ha~/~\Hb ~ ratio implies that the emission is thin in general, and is in agreement with our previous 
findings for Lupus sources \citep[see Figs. 16 and 17 in][]{frasca17}. In that paper we showed that most 
subluminous sources display large values of the Balmer decrement. Also, the trend of the Balmer decrements 
in \gqc ~shows the Type~1 form as defined in \citet[][]{antoniucci17}. For \gqc ~$\log$(H9/H4)$=-0.93$, 
which is consistent with the definition $\log$(H9/H4)$<-0.8$ for Type 1 in \citet[][]{antoniucci17}. 
The prototype of this class is the edge-on YSO  Par-Lup3-4, consistent with the edge-on geometry of 
the \gqc ~disc.

\section{Rejecting the hypothesis of a chance projection}
\label{chanceprojection}

The confirmed YSOs \citep[][and references therein]{alcala17, frasca17} in Lupus~I allowed us to define 
a range of Gaia DR2 values for $\pi$, $\mu_{\alpha}\cos{\delta}$, and $\mu_{\delta}$. However, an important constraint 
to consider here is that the spatial distribution of the Lupus~I members is not uniform, but follows 
the large-scale distribution of the cloud filament. This is more evident for the youngest YSOs 
(the Class-0 and I proto-stars) detected with the Herschel satellite \citep[][]{benedettini18}, which 
are more concentrated along the filament than the Class-II and III YSOs. We then searched for objects
lying in the same parameter space through Gaia DR2 in the Lupus~I region, and over an area of 
$\sim1.0^\circ\times6.4^\circ$ over the Lupus~I cloud filament.
However, our final Gaia DR2 parameter space is slightly wider than what we found for the actual members. 
This choice was made for two reasons: to define the sample of candidates sharing similar properties 
as those of the Class-II and Class-III bona-fide members, and to include \gqc ~in the parameter 
space despite it lying outside of parallax range (see discussion in Appendix~\ref{Appstelpar}). 

\begin{figure}[h]
 \resizebox{0.95\hsize}{!}{\includegraphics[bb=0 0 730 620]{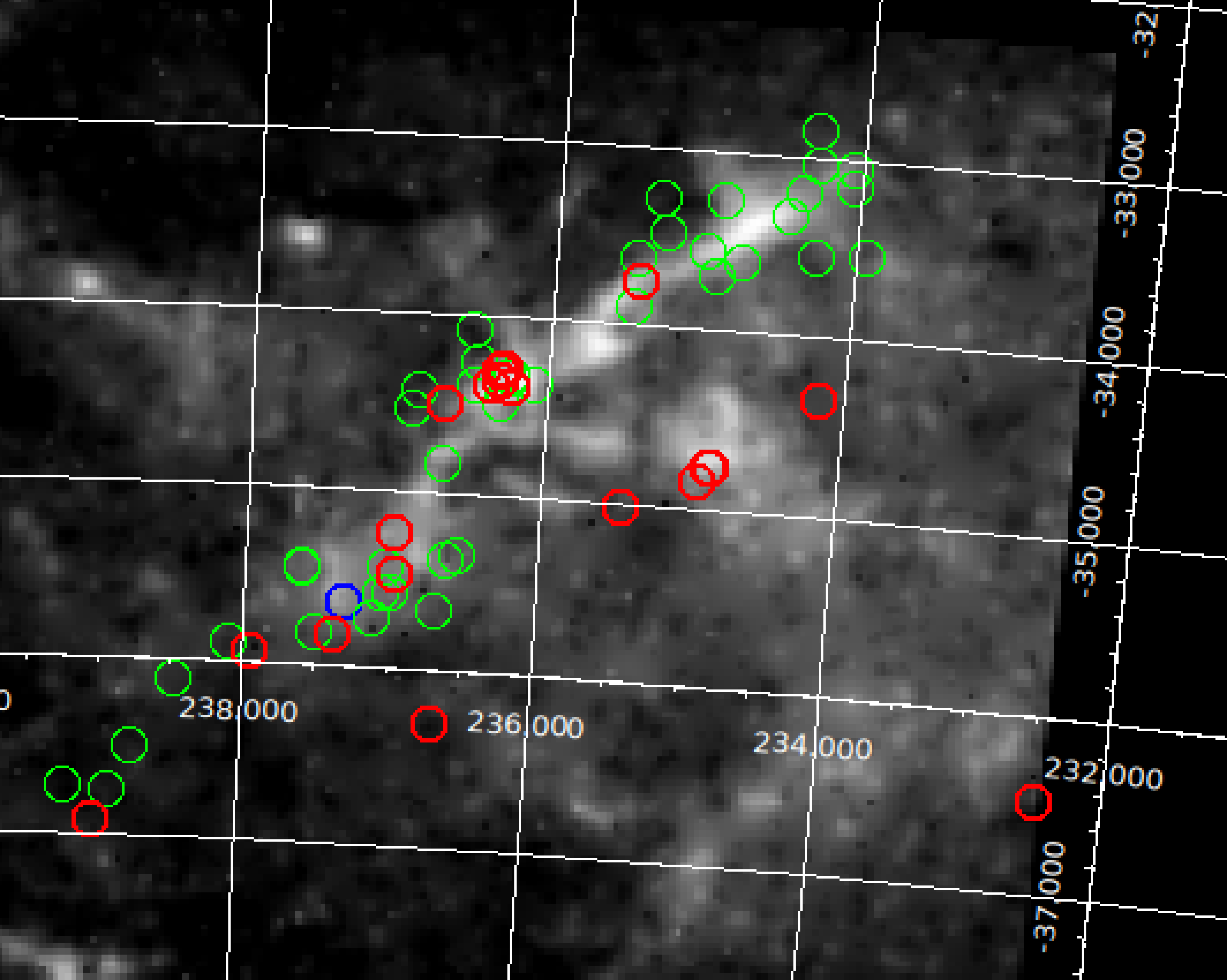}}
\caption{ Spatial distribution of the selected candidate members from Gaia DR2 (green circles) and the 
        previously known YSOs (red circles) overplotted on the extinction map of the
        Lupus~I filament. The dark blue circle shows the position of the \gq ~system.
        North is up and East to the left.
         \label{spadistr}}
\end{figure}

\begin{figure}[h]
\resizebox{1.3\hsize}{!}{\includegraphics[bb=20 0 880 550]{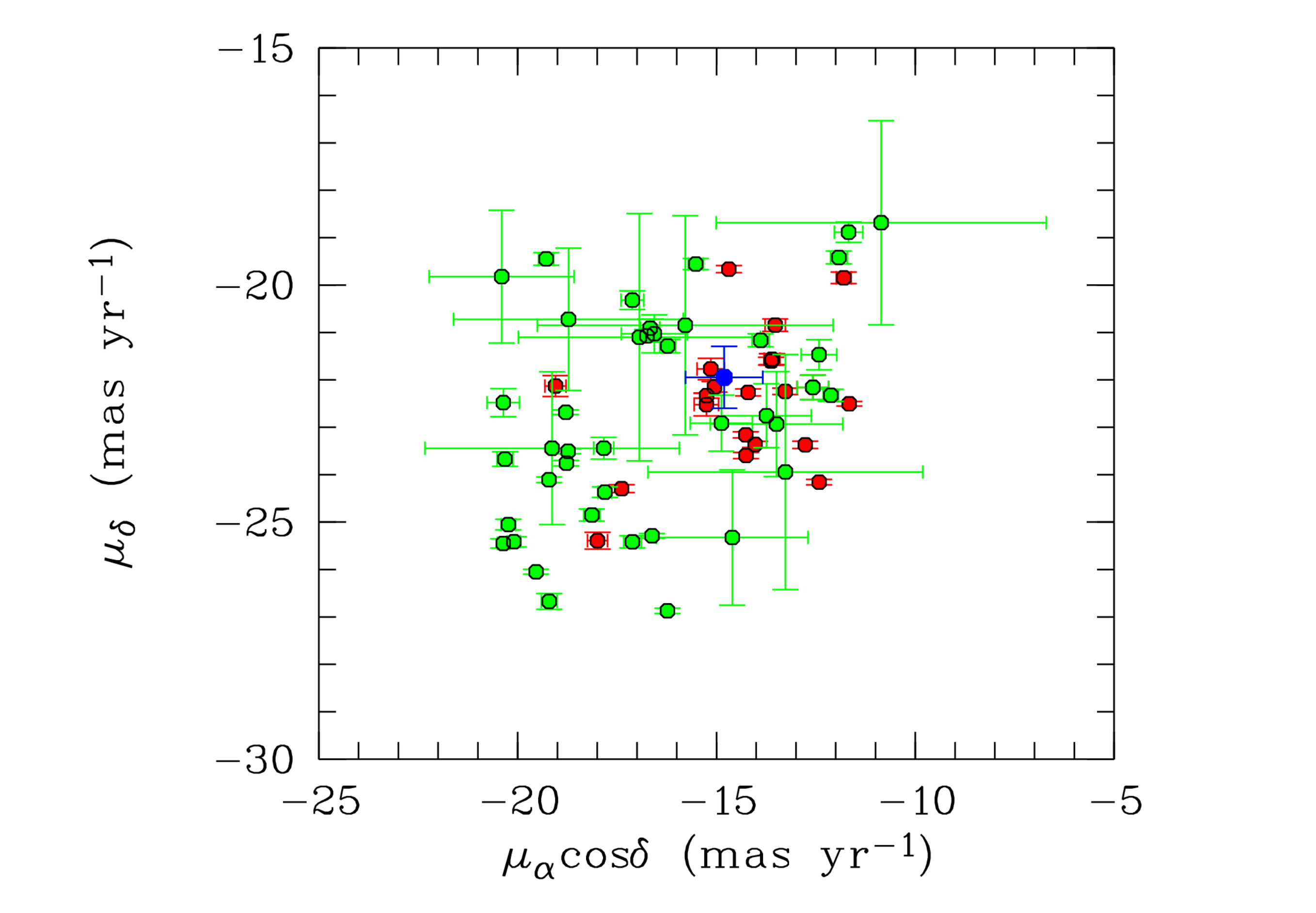}}
\caption{Scatter plot of the proper motions of the candidates (green dots) and the previously known
         YSOs in Lupus~I. The position of \gqc ~is indicated by the dark blue dot. Errors are overplotted. 
         \label{mua_mud}}
\end{figure}

The result is the following: for 5$<\pi$(mas)$<$8, $-21<\mu_{\alpha}\cos{\delta}$(mas/yr)$<-10$, 
$-27<\mu_{\delta}$(mas/yr)$<-18$, we found 55 objects in the selected region (see Figure~\ref{spadistr}).
These include 42 candidates belonging to Gaia DR2 and 13 previously known members lying on the Lupus~I filament. 
This results in a linear density of 0.0024\,objects/arcsec or an average object separation of 418\,arcsec, 
which is 26 times the observed separation between \gqa ~and \gqc. The low chances of finding an object accidentally 
in the vicinity of \gq ~supports the high probability of a physical association to the Lupus~I cloud, enhancing 
the chances of \gqa-B ~and \gqc ~ being physically bound. This is a robust result as the estimated linear density 
is only an upper limit, as several of the selected candidates are expected to be Sco-Cen members or objects 
unrelated to Lupus~I. In fact, some of the selected candidates appear rather scattered on the 
$\mu_{\alpha}\cos{\delta}$ versus $\mu_{\delta}$ scatter plot in comparison with the previously known YSOs 
(see Figure~\ref{mua_mud}).

It is also worth noting that the typical linear density of the proto-stars along the Lupus~I cloud filament 
\citep[$\sim$100/degree][]{benedettini18} is consistent with an average separation of a few tens of arcsec, 
implying a few thousand astronomical units at the distance of the Lupus clouds, and on the order of magnitude of the projected 
separation between \gqa ~and \gqc. Turbulent fragmentation is expected to occur on a similar scale 
to the thickness of the filament, which is also on the order of a few thousand astronomical units. All this would 
support the scenario of turbulent fragmentation along a filament for the formation of the wide binary. 
We therefore conclude that if \gqc ~formed within the same filament as \gqa, then they have a common origin, 
even though they might be gravitationally unbound.

\end{appendix}

\end{document}